\documentclass[aps,prd,twocolumn]{revtex4-2}
\usepackage{amsmath,amssymb,hyperref,braket,bm,graphicx}
\usepackage{tikz}
\usepackage{float}
\usepackage[none]{hyphenat}
\usetikzlibrary{positioning}
\usetikzlibrary{arrows}
\usetikzlibrary{shapes.geometric}
\usetikzlibrary{calc}
\usetikzlibrary{decorations.pathmorphing}

\begin{document}
\title{Relational Emergent Time for Quantum System: A Multi-Observer, Gravitational, and Cosmological Framework}
\author{Amir Hossein Ghasemi}
\affiliation{Independent Researcher}

\begin{abstract}
    We present a relational framework in which temporal structure is not fundamental but emerges from correlations within a globally stationary quantum state. Each subsystem includes an internal clock, and conditional states evolve effectively with respect to these internal readings. The construction naturally extends to relativistic motion, gravitational redshift, and cosmological expansion, leading to a unified emergent-time functional valid across diverse physical regimes. The theory reproduces classical time dilation, predicts correlation-dependent deviations from standard evolution, and suggests that non-interacting or massless particles exhibit negligible internal time. These consequences open directions for conceptual and experimental investigations in the foundations of temporal physics, from multi-clock quantum systems to precision metrology and cosmological settings. In particular, the framework suggests measurable deviation from standard quantum evolution for highly entangled systems and predicts negligible internal time for massless particles.
\end{abstract}

\maketitle

\section{introduction}
    The treatment of time in physics remains inconsistent: quantum theory uses an external classical parameter, whereas relativity ties temporal flow to geometry. We propose a theory where time is intrinsically relational, arising from quantum correlations rather than existing as a universal variable. In this approach, each subsystem possesses an internal clock degree of freedom whose correlations with the global state define an observer-specific emergent time, inspired by the Page-Wootters framework \cite{PageWootters}. This mechanism yields a multi-observer temporal structure compatible with relativistic dilation, gravitational redshift, and cosmological expansion \cite{RovelliTime,RovelliRelationalQM}. The theory provides a single formalism capable of describing time from the perspective of quantum mechanic \cite{UnruhWald,Kuchar}.
    Unlike classical approaches, this framework removes the need for an external time parameter, offering a fully relational description of temporal evolution.

\section{Emergent Time Framework}
To clarify the conceptual basis of the model, we adopt a set of assumptions inspired by the Page-Wootters framework:
ASSUMPTIONS:
\begin{enumerate}
   \item The universe as a whole is described by a globally timeless quantum state, This implies that there is no fundamental trace of the passage of time at the level of the cosmic background \cite{PageWootters,RovelliTime}.

   \item Each subsystem, such as an observer or any localized physical system defines its own emergent time parameter through internal correlation between the Clock degrees of freedom and the System degrees of freedom. Operationally this relies on two quantum states; the Global Clock state and a Subsystem state. \cite{RovelliRelationalQM}

   \item The rate at which emergent time flows may differ between subsystems, in in analogy with relativistic time dilation in general relativity. This variation arises from differences in their correlation with the global clock sector. \cite{MisnerThorneWheeler}
\end{enumerate}

\section{Basic structure of the model:}
This section outlines the quantum-mechanical structure underlying the emergent-time framework. We first introduce the Hilbert-space decomposition and the role of correlations between clock and subsystem degrees of freedom. Then we show how an effective Schr\"odinger evolution arises for a localized subsystem within this composition.
We need to consider two main rules about the emergent-time model:
First, in the emergent-time framework: the universe as a whole is described by a global quantum state:\, $\ket{\psi_{\mathrm{universe}}}$ \cite{PageWootters}.
The clock subsystem defines time relationally because its internal states can be ordered and serve as labels for correlations with other subsystems. Thus, the temporal parameter is identified as: $t = qo$.
And the second, temporal evolution is relational between subsystem: Since the global state of the universe is static, temporal evolution cannot arise from a global, absolute time. Instead, evolution emerges from conditional correlations between the clock subsystem and the remaining subsystems. Formally, the universe can be written in a relational composition such as: $\ket{\psi_{\mathrm{universe}}} = \sum_t\ket{t}_o \otimes \ket{\psi_{s}(t)}$, the relational nature of time follows from the constraint:
$H_{O} + H_{S} = o$.

\subsection{1. Mathematical framework:}
In this model, the universe is treated as a collection of distinguishable quantum subsystems. Each subsystem-whether an observer, a clock, or any localized physical degree of freedom-is assigned its own Hilbert space ${\mathcal{H}_i}$. The full Hilbert space is therefore the tensor-product structure,
\begin{equation}
    \ket\phi \in \mathcal{H} =
    \mathcal{H}_1 \otimes \mathcal{H}_2
    \otimes \cdots \otimes \mathcal{H}_c 
\end{equation}
And for the total Hamiltonian:
\begin{equation}
    \hat{H}_{\text{total}} = \sum_{i=1}^{N} \hat{H}_i \qquad \hat{H}_{\text{total}} \ket{\phi} = 0
\end{equation}
To extract physical time evolution from a globally static state, one isolates a particular
subsystem to serve as an internal clock. When the global state is expanded in a basis of distinguishable clock states $\ket{t_i}_C$, it takes the correlated from:
\begin{equation}
    \psi\ket{(t)} = \sum_{i,j} c_{ij} \, |
    t_i\rangle_c \otimes |\phi_j(t)\rangle_s
\end{equation}
This projection procedure is what leads to an effective Schr\"odinger equation for the subsystem, with the parameter $t$ emerging from the internal correlations rather than existing as fundamental quantity. \cite{PageWootters,OreshkovCerf}

In order to derive an effective Schr\"odinger equation for a subsystem from a globally stationary state, we follow the approach originally proposed by Page and Wootters in this framework, one isolates a particular subsystem to serve as an internal clock, and the global state is expanded in a basis of distinguishable clock states. This procedure allows time to emerges relationally, conditioned on the correlations between the and the subsystem of interest. If the entanglement is appropriately configured, the evolution of each subsystem then follows the standard Schr\"odinger dynamics, and the effective evolution equation describes the subsystems dynamics parametrically in terms of this internally defined temporal variable, without invoking an external, absolute time. While our presentation is self-contained, this method builds upon the principles of relational time as formalized in the Page and Wootters framework \cite{PageWootters}. 
The effective Schr\"odinger equation for the subsystem then reads:
\begin{equation}
    i\hbar \frac{\partial}{\partial t} \ket{\phi(t)}
    = {H}_S \ket{\phi(t)}
\end{equation}

when the subsystems are in motion or reside in the discrete gravitational potential, their respective time flows differ, consist with the principles of special and general relativity. 
As an illustrative configuration, consider three observers (A,B and C), each forming an independent subsystem, under the following assumptions:
\begin{enumerate}
    \item Each observer defines an independent subsystem.
    \item Each subsystem possesses its  own internal clock.
    \item All clocks are entangled with a common global clock source.
    \item Each subsystem occupies a distinct spatial position.
\end{enumerate}

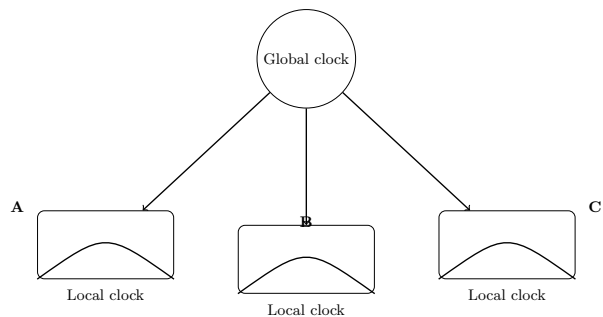
\begin{figure}[H]
\centering
\resizebox{0.45\textwidth}{!}{
\begin{tikzpicture}[
    clock/.style={draw, rounded corners, minimum width=2.8cm, minimum height=1.4cm},
    arrow/.style={->, thick},
    wave/.style={thick}
]

% Global clock
\node[circle, draw, minimum size=2.0cm] (G) {Global clock};

% Local clocks
\node[clock, below left=2.4cm and 2.0cm of G] (A) {};
\node[clock, below=2.4cm of G] (B) {};
\node[clock, below right=2.4cm and 2.0cm of G] (C) {};

% Labels
\node[above left=-0.15cm and 0.15cm of A] {\textbf{A}};
\node[above= -0.15cm of B] {\textbf{B}};
\node[above right=-0.15cm and 0.15cm of C] {\textbf{C}};

\node[below=0.1cm of A] {Local clock};
\node[below=0.1cm of B] {Local clock};
\node[below=0.1cm of C] {Local clock};

% Define wave paths with coordinates
\path (A.south west) coordinate (Aw) (A.south east) coordinate (Ae) (A.south) coordinate (Am);
\path (B.south west) coordinate (Bw) (B.south east) coordinate (Be) (B.south) coordinate (Bm);
\path (C.south west) coordinate (Cw) (C.south east) coordinate (Ce) (C.south) coordinate (Cm);

% Draw waves
\draw[thick] (Aw) .. controls ($(Am)+(0,1.0)$) .. (Ae);
\draw[thick] (Bw) .. controls ($(Bm)+(0,1.0)$) .. (Be);
\draw[thick] (Cw) .. controls ($(Cm)+(0,1.0)$) .. (Ce);

% Arrows
\draw[arrow] (G) -- (A);
\draw[arrow] (G) -- (B);
\draw[arrow] (G) -- (C);

\end{tikzpicture}}

\caption{
Schematic of the emergent-time framework.  
A stationary global clock sector entangles with three local clock subsystems (A, B, C), generating distinct relational time parameters for each observer.
}
\label{fig:emergent_time}
\end{figure}

Under these conditions, the global state can be written as:
\begin{equation}
    \ket\Phi = |t\rangle_c \otimes |
    S_A(t_A)\rangle \otimes |S_B(t_B)\rangle 
    \otimes |S_C(t_C)\rangle
\end{equation}
With assumption $t_A \neq t_B t_C$ and $\quad \frac{dt_A}
{dt} \neq \frac{dt_c}{dt}$.
To illustrate the relational structure of emergent time in our framework, we consider a setup with one global clock entangled with multiple local clock subsystems. Figure~\ref{fig:emergent_time} schematically depicts how the global clock correlates with three local (A,B,C), giving rise to distinct temporal parameters for each observer. This visualization emphasizes the core idea that time in each subsystem emerges from internal correlations rather than existing as a fundamental background quantity.

This diagram illustrates three key features:

(1) Time is not a global background parameter: the universe is described by a stationary global clock sector, while each subsystem (A, B, C) defines its own local time through internal correlations.

(2) Differences in the entanglement structure between each local clock and the global clock lead to distinct relational time flows, producing quantum-induced time dilation effects between subsystems.

(3) Standard Schrödinger dynamics for each subsystem emerges effectively from projecting the global stationary state onto local degrees of freedom, rather than being a fundamental global evolution.

This example provides an intuitive visualization of how local time parameters.

\subsection{2. Mathematical Framework: Emergent Schr\"odinger Dynamics:}
In this section, we formalize how time can emerge for a subsystem from a globally stationary state. As we said, the universe is modeled as a composite system consisting of a clock and a physical subsystem. By projecting the global entanglement state onto a definite clock reading, we can define conditional subsystem states that evolve according to an effective Schr\"odinger equation. This procedure allows local time to appear relationally, arising from correlations between the clock and the subsystem, without invoking any external absolute time.
In the second stage of the framework, the universe is modeled as a composite system consisting of a clock and physical subsystem.
The global entangled state is represented as:
\begin{equation}
    \ket{\phi} = \sum_t \ket{t}_c \otimes \ket{s_t}_S ,
    \label{eq:entangled_state}
\end{equation}

Where ${\ket{t}_C}$ denotes the orthonormal clock basis and $\ket{s_t}_S$ represents the corresponding conditional states of the system.
The total state satisfies the stationarity (Hamiltonian constraint) condition:
\begin{equation}
    \hat{H}_{\mathrm{tot}}\ket{\phi} = 0
    \qquad
    \hat{H}_{\mathrm{tot}} = \hat{H}_C + \hat{H}_S + \hat{H}_{\mathrm{int}}
\end{equation}
Projecting the total state onto a definite clock reading $t$, the conditional subsystem state implies as $\ket{\psi_S(t)} \propto \bra{t}_C \ket{\phi}$, Which evolves according to an effective Schrodinger equation with respect to the emergent parameter $t$ \cite{GambiniPullin,PageWootters},
\begin{equation}
    i\hbar, \frac{\partial}{\partial t}
    \ket{\psi_S(t)} = \hat{H}_S\
    \ket{\psi_S(t)}
    \label{eq:schrodinger}
\end{equation}
This projection \ref{eq:schrodinger}, procedure leads to an effective Schr\"odinger equation for the subsystem, as originally shown in the Page-Wootters framework \cite{PageWootters}
Here Equation \ref{eq:schrodinger}, $t$ does not correspond to an external absolute time, but instead emerges from correlation between the clock and the system.
Consequently, each observer associated with a distinct subsystem experiences a local, emergent time evolution governed by its own effective Hamiltonian. 
Under the stationary constraint, the total Hamiltonian of the composite configuration reduces to:
\begin{equation}
    \hat{H}_{\mathrm{tot}} = \hat{H}_C + \hat{H}_S
    \qquad
    \hat{H}_{\mathrm{tot}}\ket{\phi} = 0
    \label{eq:stationary_condition}
\end{equation}

Equation \ref{eq:stationary_condition}, Ensures that the global state $\ket{\phi}$ remains timeless, while local dynamics arise relative to each subsystems clock degree of freedom.
With considering the stationary condition $\left(\hat{H}_C + \hat{H}_S\right)\ket{\phi} = 0$ and taking the inner product with $\bra{t}_C$ yields:
 \begin{equation}
     i\hbar\,\frac{\partial}{\partial t}\ket{\psi_S(t)}
     \label{eq:emergent_schrodinger}
 \end{equation}
 The resulting effective Schr\"odinger equation describes the dynamics of the subsystem with respect to the emergent internal time. This construction highlights two important points: first, the global wave function remains timeless, and second, the local dynamics experienced by each subsystem are determined entirely by its entanglement with the internal clock. For multiple obervers or subsystems, the same procedure generalizes naturally, producing locally emergent times for each subsystem while preserving consistency with the stationary global state.
This represents the emergent Schrodinger equation for the subsystem.
This construction implies that
\begin{enumerate}
    \item The total wave function of the universe is globally timeless.
    \item Time appears as an emergent parameter resulting from the entanglement between the clock and the system.
\end{enumerate}
For an ensemble of $n$ observers, this process generalizes naturally, giving rise to multiple locally emergent times, each associated with a distinct subsystem.
In the framework of emergent Schr\"odinger dynamics \cite{PageWootters}, time emerges relationally from the entanglement between the clock and the subsystem. To illustrate how this entanglement effects the local dynamics, we introduce two key quantities:
\begin{itemize}
    \item E: the amount of entanglement between the clock and the subsystem.
    \item C: the local coherence (or independence) of the subsystem.
\end{itemize}
A commonly used and physically reasonable relation between these quantities is given by $C(E) = C_{0} \, e^{-k E}, \quad k>0$. 
To provide a clearer understanding of this relationship, we present the plot below, which directly visualizes how local coherence decreases as entanglement increases according to the above equation:
Illustrative Example: Three Observers with Distinct Emergent Times: To make the structure concrete, consider three subsystems-observer that each entangled to different degree with the internal clock. Although the global state is $\psi _{\text{tot}}$. Let the intrinsic, observer-independent evolution be governed by:
\begin{equation}
    \frac{d}{d\tau}\psi_{\text{tot}}
    (\tau)=\mathcal{G}
    \psi_{\text{tot}}(\tau),
\end{equation}
Each observer introduces a local time through the relation $\frac{d\tau}{dt_i}=f_i(E_i)$. This immediately produces an emergent Schr\"odinger equation for each observer:
\begin{equation}
    i\hbar\frac{\partial}{\partial t_i}\psi_i(t_i)=H_i \, \psi_i(t_i)
\end{equation}
Thus, the same intrinsic generator appears slower or faster depending on the entanglement strength.
Relation to Local Coherence: Each subsystem also carries a local coherence, representing the degree to which it behaves as an autonomous quantum system. Motivate by the dynamical structure gives $C_i(E_i)=C_0 \, e^{-k \, E_i}$ that we can conclude that:
\begin{enumerate}
    \item Observer A: weak entanglement, nearly maximal coherence, fast and almost standard Schr\"odinger evolution.
    \item Observer B: moderate entanglement, partially reduced coherence, slower perceived dynamics.
    \item Observer C: strong entanglement, heavily suppressed coherence, evolution appears strongly "dephased".
\end{enumerate}
The three observers therefore assign different dynamical timescales and different coherence decays to the same underlying global process.
Decoherence-like behavior: Although the global evolution is unitary, the local subsystem dynamics exhibit a decoherence-like effect purely due to entanglement with the clock.
To see this explicity, consider an initial local superposition for each observer:
\begin{equation}
    \psi_i(0)=\alpha \ket{0}+\beta\ket{1} 
\end{equation}  
\begin{equation}
    \psi_i(t_i)=\alpha e^{-i\omega t_i}\ket{0}+\beta e^{-i\omega t_i} e^{-kE_i} \, \ket{1}
\end{equation}
The phase evolution is the same for all observers, but the relative amplitude is attenuated by the factor.
The effective reduced density matrix becomes:
\begin{equation}
    \rho_i(t_i)=
    \begin{pmatrix}
        |\alpha|^2 & \alpha|\beta^* e^{-kE_i} \\
        \alpha^*\beta e^{-kE_i} & |\beta|^2
    \end{pmatrix}
\end{equation}
The off-diagonal terms shrink as grows.
This reproduces as clean, observer-dependent decoherence pattern, even though the global state remains coherent and timeless.
This three-observer model shows concretely that:
\begin{enumerate}
    \item Time emerges relationally from entanglement with the internal clock.
    \item Different observers acquire different effective Schr\"odinger equations, determined only by their entanglement strength.
    \item local coherence decays exponentially with entanglement, matching the expression.
    \item Apparent decoherence is not fundamental, but a natural consequence of tracing out the clock-subsystem correlations.
\end{enumerate}
This aligns directly with the plotted function and justifies the use of as a measure of local quantum autonomy.
\begin{figure}[H]
\centering
\includegraphics[width=0.4\textwidth]{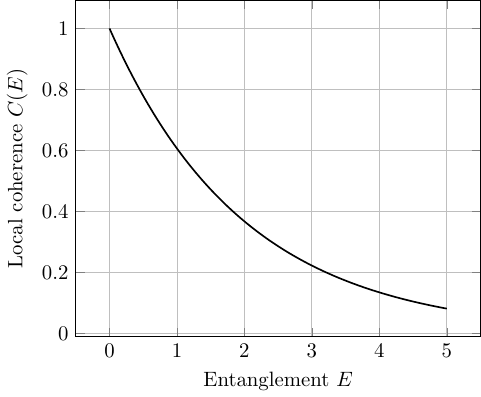}
\caption{Figure: Relationship between entanglement and local coherence in an emergent Schr\"odinger dynamics framework. The horizontal axis represents the entanglement between the clock and the subsystem, while the vertical axis shows the local coherence of the subsystem. According to the standard model, local coherence decreases exponentially with increasing entanglement. This plot illustrates the monotonic decay of local coherence as entanglement grows.}
\end{figure}

\section{extension to multiple observer(N system-model)}
The framework naturally generalizes to multiple subsystems correlated through a single global clock:
\begin{equation}
    \ket{\phi} = \sum_t \ket{t}_C \otimes \ket{s_1(t)}_{s_t} \otimes \ket{s_2(t)}_{S_2} \otimes \cdots \otimes \ket{s_n(t)}_{s_n}
    \label{eq:multiobserver_state}
\end{equation}

Where $\ket{s_i(t)}_{s_i}$ denotes the conditional states of the $i$ subsystem relative to the same clock reading $t$.
Each subsystem defines its own emergent emergent time parameter through local correlations with the clock, leading observer-dependent temporal evolution \cite{RovelliRelationalQM,GambiniPullin}
Within this framework, time differences arise naturally among observers due to distinct local Hamiltonians $\hat{H}_S$ \cite{WignerClock} or variations in spacetime condition, thereby providing a consistent bridge between quantum correlation and relativistic time dilation.
To incorporate relativistic effects, we extend the model to observers located in distinct gravitational potentials or moving at different velocities.
For instance, consider three systems: one on Earth, one on Mars, and one aboard a spacecraft in motion. 
According to special relativity \cite{MisnerThorneWheeler}, each observer experiences a different proper time given by;
\begin{equation}
    dt^{(i)} = \gamma^{(i)} dt, \qquad \gamma^{(i)} = \frac{1}{\sqrt{1 - \frac{v_i^2}{c^2}}}
\end{equation}

In general relativity \cite{MisnerThorneWheeler,WaldGR}, gravitational time dilation near a mass $M$ is expressed as;
\begin{equation}
    d\tau^2 = \left(1 - \frac{2MG}{rc^2}\right) dt^2
\end{equation}

Thus, although the global wavefunction of the universes is defined at a common coordinate time $t$ each observer perceives a distinct temporal rate due to relative motion and gravitational potential.
For each observer $i$, the local emergent time evolution of the subsystem follows a Schrodinger-like equation;
\begin{equation}
    i\hbar ,\frac{d}{dt_{(i)}}\ket{s_{t_{(i)}}} = \hat{H}_{S}^{(i)} \ket{s_{t_{(i)}}}
\end{equation}

Hence, time for each subsystem emerges from quantum entanglement, while  its rate depends on the underlying spacetime geometry. In that case, we can examine time from a different frame than the relativity frame.
In the extended model, the universe can be visualized as a multidimensional "hypercube" of all possible configurations \cite{WignerClock}.
Each observer perceives only a particular cross-section of this hypercube a temporal slice-a local temporal slice- which is interpreted as the flow of time; 
\begin{equation}
    \ket{\phi} = \sum_{t} \ket{t}_{c} \otimes \ket{s_{1}(t)} \otimes \ket{s_{2}(t)} \otimes \cdots \otimes \ket{s_{n}(t)} 
\end{equation}
Because each observer experiences a distinct relative time rate-due to both relative motion and gravitational potential-their effective time variable becomes an observer-dependent emergent parameter.
Following the Page-Wootters mechanism and its extension, the conditional state of subsystem relative to clock is $\ket{\psi_i(\tau_i)}=\bra{\tau_i}_C \ket{\phi}$ .
Under this condition, the observer-dependent state satisfies an emergent Schr\"odinger equation:
\begin{equation}
    i\hbar\frac{\partial}{\partial\tau_i}\ket{\psi_i(\tau_i)}= H_{S}\ket{\psi_i(\tau_i)}
\end{equation}
The relation between global time and local emergent time is set by relativity:
\begin{equation}
    \mathrm{d}\tau_i = \sqrt{1 - \frac{v_i^2}{c^2}-\frac{2GM}{r_i c^2}}\, \mathrm{d}t.
\end{equation}
Thus, each subsystem acquires local time generated by entanglement with its internal clock, while the rate of this emergent time depends on the underlying spacetime geometry.
The perceived motion of the universe is a therefor a manifestation of observers experiencing distinct temporal slices of a globally static, entangled state.
For an observer, the local time element is given by;
\begin{equation}
    dt^{(i)} = \gamma^{(i)} \, dt, \qquad
    \gamma^{(i)} = \frac{1}{\sqrt{1 - \frac{v_i^2}{c^2}}}
\end{equation}
This shows that time appears as a relative and emergent quantity resulting from entanglement, with in rate governed by the local curvature of spacetime.
In this representation, the universe corresponds to a \textit{hypercube} containing all possible physical states.
To formalism the idea of a globally static but internally relational universe, we model the total configuration space as a high-dimensional hyper-cube (or hyper-lattice) in which every vertex corresponds to a possible complete configuration of all subsystems. The global wavefunction assigns amplitudes to these configurations.
In this representation:
each axis of the hyper-cube corresponds to a degree of freedom and the second, each vertex is full configuration of the universe.
the full global state is stationary $H_{\mathrm{tot}}\ket{\phi}=0$ .
A specific observer, however, only has access to a one-dimensional slice of this hyper-cube-selected through their internal clock state. This slice is interpreted as the observers perceived temporal evolution.
Therefore, what is experienced as "flow of time" is simply the traversal of successive conditional slice of an otherwise static entangled global state.
\begin{figure}[H]
\centering
\includegraphics[width=0.43\textwidth]{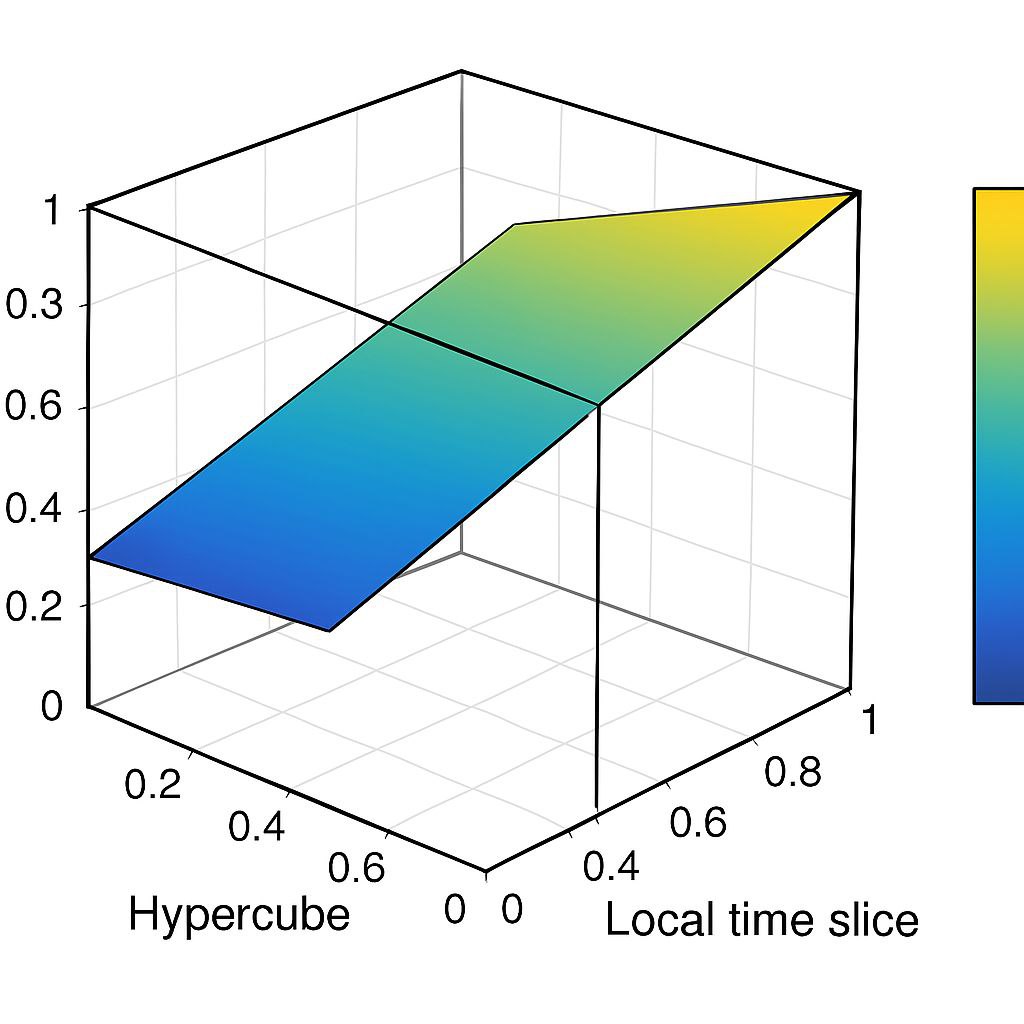}
\caption{Figure: Schematic representation of the global configuration-space hyper-cube and emergent observer-dependent time slices. The full hyper-cube encodes all possible configurations of the universe in a stationary entangled state. Each observer accesses only a conditional one-dimensional slice selected by their internal clock state, which is perceived as a continuous temporal evolution. Differences in time-flow arise from entanglement structure and relativistic time dilation.} 
\end{figure}

Each observer accesses a single cross-sectional slice, thereby perceiving the illusion of continuous temporal evolution within an otherwise static global configuration.
\subsection{Observers With Strong and Weak Entanglement}
A photon has no internal degrees of freedom capable of forming a clock that can correlated with external systems \cite{CastroRuiz,SaleckerWigner}. Thus, its conditional states do not define a meaningful internal time parameter:
\begin{equation}
    \ket{\psi_{\gamma}(\tau)} \, \text{is undefined} \qquad \mathrm{d}\tau_\gamma = 0 
\end{equation}
In contrast, macroscopic observer such as human are strongly entangled with their environment. Their internal degrees of freedom act as stable clocks, giving rise to a well-defined emergent time variable.
The corresponding emergent Schr\"odinger equation for a human-level observer becomes:
\begin{equation}
    i\hbar\frac{\partial}{\partial\tau_{\mathrm{human}}} \ket{\psi_{\mathrm{human}} = H_{S} \ket{\psi_{\mathrm{human}}(\tau_{\mathrm{human}})}}
\end{equation}
The impossibility of accelerating a human to speed of light is consistent with the disappearance of proper time in the limit $\lim_{v\to c} \mathrm{d} \tau_{\mathrm{human}} = 0$ .
\section{emergent time formalism}
The quantum state of the total configuration can be written as:
\begin{equation}
\ket{\psi_{s}} = \sum_n \, C_{N} \ket{t_{n}} \ket{s_{N}}
\end{equation}

Where the coefficients encode correlations between the clock degrees of freedom and the system state.
This composite structure implies the existence of a shared temporal reference emerging from the dynamical entanglement among subsystems.
The total Hamiltonian of the closed universe satisfies the stationary (constraint) condition:
\begin{equation}
H_{\text{total}} = H_{1} + H_{2} + \dots + H_{n} + H_{\text{clock}} = 0
\end{equation}

This constraint enforces global timelessness: three exists no external or absolute time parameter for the universe as a whole.
Instead, temporal evolution emerges locally within subsystems through their correlation with the clock degrees of freedom.
Hance, "time" in this formulation is not a background variable but a relational quantity, arising directly from the entanglement between the clock and the physical system.

\section{Verification of The Temporal Flow Model}
\subsection{A. Structural Examination in Gravitational Contexts}
we consider two stationary observers \(A\) and \(B\), that located at fixed Schwarzschild radii $r_{A}$ and $r_{B}$ respectively, outside the event horizon of a spherically symmetric mass $\mathrm{M}$. In Schwarzschild coordinates (and the natural units $c = 1$) \cite{WaldGR,MisnerThorneWheeler} the relation between the coordinate time $t$ measured at spatial infinity and the proper time $\tau$ measured by a static observer at radius $r$ is:
\begin{equation}
    d\tau = \sqrt{1 - \frac{2GM}{r}},\, \qquad \tau_{B}(t) = \int \sqrt{1 - \frac{2GM}{r_{B}}}\, dt
\end{equation}
In our emergent-time construction we introduce the quantity $t^{\mathrm{emergent}}_{B}$ defined by :
\begin{equation}
    t_{B}^{\text{(emergent)}} = \int \! M \, \sqrt{ 1 - \frac{2GM}{r_B}} \, dt \;=\; M\,\tau_{B}(t)
\end{equation}
A concise interpretation of the expressions above is as follows. The factor $\sqrt{1 - \frac{2GM}{r_{B}}}$ is the standard Schwarzschild gravitational redshift (time-dilation) factor for a static observer located at redius $r_{B}$ . The integral $\tau_{B} = \int \sqrt{1 - \frac{2GM}{r_B}},dt$ represents the general-relativistic proper time accumulated by observer $B$. Finally, the relation $t_{B}^{\mathrm{emergnet}} = M, \tau_B$ identifies the emrgent-time variable with the $GR$proper time up to a model-dependent normalization constant $M$.
When $M$ is fixed so that the emergent variable carries the same units as the coordinate-time parameter (or any chosen microscopic clock), the two notions become operationally equivalent.
The relation above relies on several implicit assumptions:
\begin{enumerate}
    \item The background geometry is Schwarzschild space-time static, spherically symmetric, and vacuum outside the mass source-and Schwazschild coordinates $(t,r,\theta,\phi)$ are used throughout.
    \item Observers $A$ and $B$ remain static $(dr = d\Omega = 0)$, meaning they are supported at radii by non-gravitational forces. The expression for $d\tau$ does not apply to freely falling observers except along their specific trajectories.
    \item Geometric units with $c = 1$ are adopted in $SI$ units, one restores $c$ by replacing $\frac{2GM}{r}$ with $\frac{2GM}{c^2r}$ and $dt$ with $c,dt$.
    \item The factor $M$ in $t^{\mathrm{emergent}}_B$ represents a normalization scale intrinsic to the emergent-time framework. It either fixes the dimensional conversion between emergen-time and $GR$ proper time or may carry physical meaning within the underlying model.
    \item effects such as back-reaction of the observers clock, or any coupling between the emergent degrees of freedom and the spacetime geometry (e.g.,modifications of $M$ or $r_B$), are neglected in this approximation. 
\end{enumerate}
Since the emergent -time expression contains the same redshift factor $\sqrt{1 - \frac{2GM}{r_B}}$, that appears in general relativity, it reproduces the correct radial dependence of gravitational time dilation for static observers near a Shhwarzschild black hole.
Thus, within the range of assumptions above, the emergent-time framework yields the same physical predictions as $GR$ regarding how clocks slow down near the event horizon. Any genuine deviation from $GR$ in this model must therefore arise either from modifications of the radial dependence, from non-local contributions to the emergent-time, or through dynamical variations of the normalization constant $M$.

\subsection{B. Gravitational Effect on Emergent Schr\"odinger Dynamics}
The subsystem maintains its state via entanglement with the global clock.
As approaches the horizon, the local temporal rate slows due to redshift.
In flat spacetime, the self-emergent Schrodinger dynamic is:
\begin{equation}
    i\hbar \frac{d}{dt}\ket{s(t)} = \hat{H}_{S}\ket{s(t)}
\end{equation}

In curved spacetime, the emergent-time parameter inherits the redshift structure of the background metric. As the subsystem maintains entanglement with the global clock degree of freedom, the rate at which the effective time parameter flows becomes locally modulated by the gravitational potential. This modification enters only through the mapping $dt\mapsto dt_{\mathrm{emergent}}$ , while the internal Hamiltonian $\hat{H}_S$ and the linearity of the Schr\"odinger evolution remain unchanged.
\begin{equation}
    i\hbar \, \frac{d}{dt_{\text{emergent}}} \, | s(t) \rangle = \hat{H}_{S} \, | s(t) \rangle\
\end{equation}
The structure  of this equation is formally identical to the standard Schr\"odinger equation,but its operational meaning is different: the "clock" that defines the time parameter is no longer an external classical variable but an emergent quantity tied to the gravitational environment. Consequently, the gravitational field does not modify the Hamiltomian term, but instead rescale the temporal parameter with the same redshift factor appearing in general relativity.  
thus, the formalism remains valid under gravitational influences, with time dilation manifesting as a reduction in emergent temporal flow.
The reduction in emergent temporal flow reflects the fact that the entanglement-induced clock ticks more slowly in regions of deeper gravitational potential. This slowdown is not imposed by hand but arises automatically from the dependence of $t_{\mathrm{emergent}}$ on the Schwarzschild redshift factor. In this sense, emergent-time dynamics encodes gravitational time dilation at the level of quantum evolution without modifying the subsystem Hamiltonian.
Taken together, these results show that the emergent-time formalism remains consistent in curved spacetime: the subsystem evolves according to an unmodified Hamiltonian, while gravitational effects manifest exclusively through the deformation of the emergent temporal parameter. This leads to a quantum-mechanical expression of gravitational time dilation that is structurally equivalent to the classical general-relativistic relation.

\subsection{C. Accelerated Expansion and the Emergent Time in FLRW Spacetime}
In order to understand how emergent time behaves in a dynamically evolving universe \cite{HartleHawking}, it is essential to analyze the formalism within a cosmological background. The FLRW spacetime provides the natural framework since any local subsystem is inevitably coupled-through entanglement-to the global cosmological clock defined by the scale factor $a(t)$.
For an expanding universe governed by the cosmological constant, the flat FLRW metric is :
\begin{equation}
    ds^{2} = - c^{2} dt^{2} + a(t)^{2} \left(dr^{2} + r^{2} d\Omega^{2} \right)
\end{equation}
Where \(\Omega\) is the scale factor.
In this setting, the local subsystem does not evolve with respect to the coordinate time $t$ , but rather relative to the effective time parameter extracted from its entanglement with the cosmic clock. The stretching of spacetime encoded in $a(t)$ directly modifies the local rate at which emergent time flows.
The emergent time of a local subsystem entangled with the cosmic clock is:
\begin{equation}
    t^{\text{emergent}}_{(i)} = \int{\frac{dt}{a(t)}}
\end{equation}
The inverse dependence on the scale factor shows that the faster the universe expands, the slower the emergent time accumulates for the subsystem. This parallels gravitational redshift, indicating that cosmic expansion acts as a "kinematic redshift" on emergent temporal flow. 
For exponential expansion, we find:
\begin{equation}
    t^{\text{emergent}}_{(i)}  = \int e^{-Ht} \, dt = -\frac{1}{H} \, e^{-Ht} + const
\end{equation}
During accelerated expansion with constant Hubble parameter $H$, the emergent time saturates instead of growing indefinitely. This saturation indicates that, from the perspective of the subsystem, the universe asymptotically approaches a "frozen-time" regime where the local dynamics become increasingly suppressed.
Despite the modification of the time parameter, the emergent Schr\"odinger equation retains its form. The only change is the replacement $t\to t^{\mathrm{emergent}}_i$ , meaning that the generator of local dynamics remains $\hat{H}_S$, but the rate at which by the evolution unfolds is rescaled by the cosmological expansion. 
The corresponding dynamical law becomes:
\begin{equation}
    i\hbar \, \frac{\partial}{\partial t \,^{\text{emergent}}_{(i)}} \, |S(t_{(i)})\rangle = \hat{H}_{S} \, | S(t_{(i)})\rangle 
\end{equation}
Therefore , cosmic expansion modulates the emergent dynamical flow in precisely the same way that gravitational potentials do. Both phenomenal reduce the effective ticking rate of the emergent clock, demonstrating that emergent time is sensitive to any mechanism-geometric or gravitational-that alters the relational structure between the subsystem and its reference clock.

\subsection{D. Consistency with Special and General Relativity}
To complete the consistency of the formalism, it is necessary to verify that the emergent-time construction remains compatible with the relativistic structure of time in both special and general relativity.
\textbf{Special Relativity:}
Consider a subsystem moving with velocity $v$ relative to the global reference clock. Its proper time $\tau$ is related to the coordinates time $t$ through the standard special-relativistic time dilation.
For a subsystem moving with velocity:
\begin{equation}
    d\tau_{i} = \sqrt{1 - \frac{v^{2}}{c^{2}}} \, dt
\end{equation}
with considering the emergent time formalism and replacing the $\tau\to t$:
\begin{equation}
    t^{\text{emergent}}_{i} = \int \sqrt{1 - \frac{v^{2}}{c^{2}}} \, dt
\end{equation}
In the emergent-time framework, the subsystem necessarily evolves with respect to its own proper time $\tau$ , since entanglement with the reference clock is defined along the subsystems world-line. Therefore, the emergent Schr\"odinger equation naturally adopts $\tau$ as its dynamical parameter, without requiring any modification of its structure.
\textbf{General Relativity:}
In curved spacetime, the proper temporal parameter experienced by a subsystem is determined by the spacetime metric along its trajectory.
In general metric:
\begin{equation}
    t^{\text{emergent}}_{i} = \int \sqrt{- g_{\mu\nu}\frac{dx^{\mu}}{dt}\frac{dx^{\nu}}{dt}} \, dt
\end{equation}
This expression shows that the emergent time coincides with the relativistic proper time up to the rescaling induced by entanglement with the global reference clock. As a result, gravitational time dilation is automatically incorporated into the emergent temporal flow. 
Therefore, both in flat and curved spacetime, the emergent time formalism aligns with the standard relativistic behavior of temporal flow. It reproduces time dilation in special relativity and general relativity without additional assumptions, demonstrating that the emergent clock behaves as a fully temporal parameter.
\section{Conceptual Framework of Emergent Time}
Time is treated here not as a fundamental external parameter, but as a relational quantity emerging from quantum correlations between physical subsystem and a global clock degree of freedom, Relational and timeless approaches to quantum dynamics have been explored from different perspective \cite{RovelliTime,GambiniMontevideo,Kuchar}. In this framework, the notion of time experienced by a subsystem arises through entanglement with a reference clock, rather than being imposed a priori as an absolute background variable.
The emergent time constructions assumes a globally constrained quantum state, in which the total Hamiltonian vanishes and physical evolution is encoded relationally. Local temporal flow  is then defined conditionally, by projecting the global state onto clock states and examining the induced evolution of the subsystem. As a result, time becomes observer-dependent and system-dependent, while remaining operationally well-defined.
This relational viewpoint naturally accommodates relativistic effects, since different observers-characterized by distinct kinematic states or gravitational environments-couple differently to the underlying clock degree of freedom. The emergent temporal rate therefore depends on both quantum correlations and space time geometry, allowing a unified treatment of quantum dynamics, gravitation, and cosmology.
\section{Unified Emergent Time in Relativistic and Cosmology Settings}
We now construct a unified expression for emergent time that simultaneously incorporates gravitational, kinematic, and cosmological effects. Consider a subsystem $i$ characterized by radial position $r_i$ , relative velocity $v$, and embedded in cosmological background with expansion rate $H$, the emergent temporal parameter associated with this subsystem is defined as:
\begin{equation}
    t_i^{\mathrm{emergent}} = \int \sqrt{1 - \frac{2GM}{r_i} - \frac{v_i^2}{c^2} - H^2 r_i^2} \, dt
    \label{eq:combined_effects}
\end{equation}
In Equation \ref{eq:combined_effects}, each contribution inside the square root has a clear physical interpretation. The first term corresponds to gravitational redshift due to spacetime curvature, The second term represents kinematic time dilation arising from relative motion, and the third term encodes the influence of accelerated cosmic expansion through the presence of a cosmological horizon.
Importantly, this expression is not introduced phenomenologically. Rather, it emerges naturally within the entanglement-based framework of time, where the local temporal rate is determined by how the subsystem couples to the global clock under geometric and dynamical constraints. The combined structure reflects the fact that all three effects-gravitational, motion, and expansion-modify the rate at which correlations between the subsystem and the clock evolve.
As the subsystem approaches strong-field regimes (e.g., near a black hole horizon), or as the cosmological expansion becomes dominant, the emergent temporal flow slows down relative to a distinct reference observer. This mirrors the predictions of general relativity, while arising here from a relational quantum-mechanical construction.
\section{Multi-Observer and N-System generalization}
The emergent-time formalism admits a natural generalization to multi observer and subsystems. Consider a collection of $n$ systems, each locally entangled with a common global clock but characterized by distinct kinematic and gravitational conditions.
\begin{figure}[H]
\centering
\includegraphics[width=0.5\textwidth]{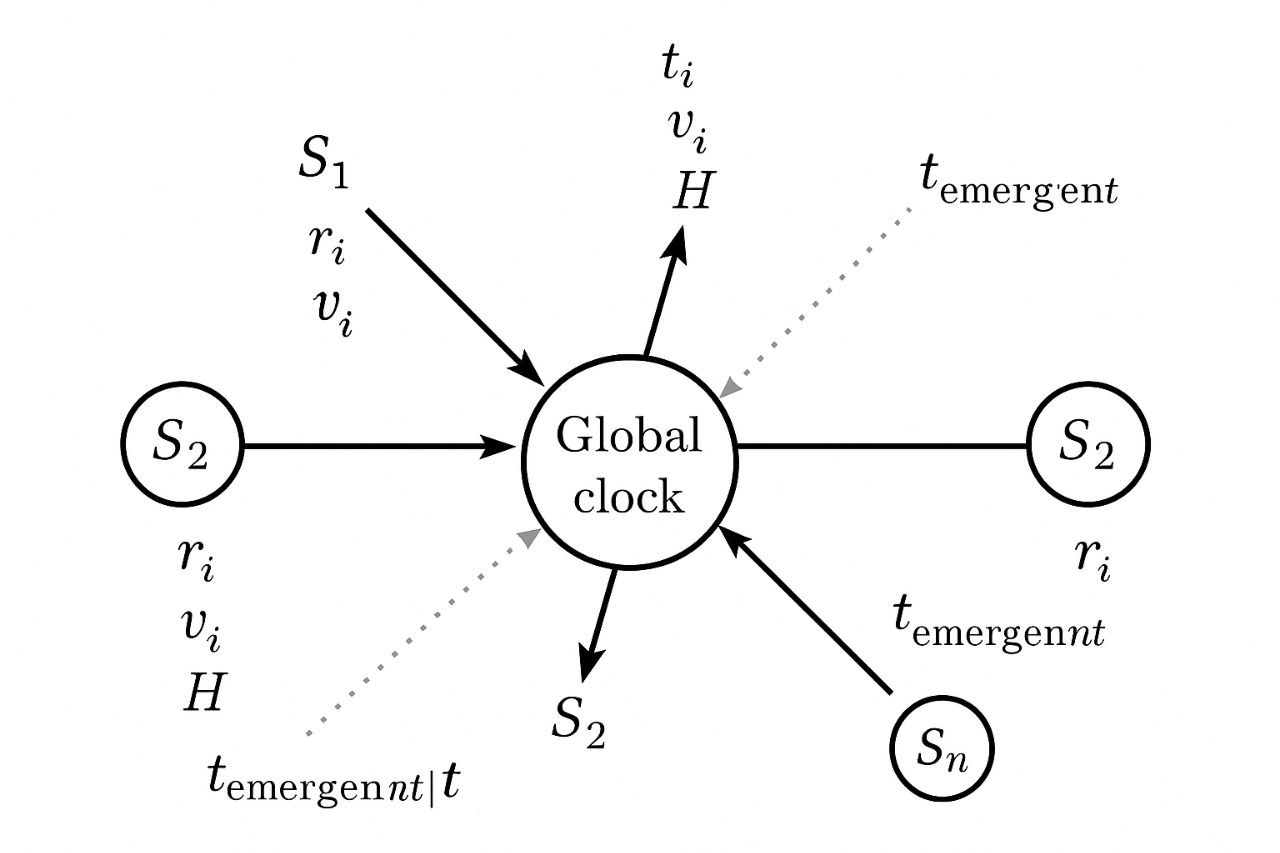}
\caption{Figure: Conceptual illustration of relational emergent time. Local subsystem $S_i$ are entangled with a global clock state, while differences in spacetime geometry and motion modify the effective rate correlation evolution. As a result, each observer experiences a distinct emergent temporal flow, reproducing relativistic time dilation within a quantum-relational framework.} 
\end{figure}
A schematic illustration of the multi-observer relational structure is shown in Fig.4.
The global quantum state may be written as:
\begin{equation}
    \ket{\phi} = \sum_t \ket{t_c} \otimes \ket{s_1(t)} \otimes \dots \otimes \ket{s_n(t)}
    \label{eq:multi_observers}
\end{equation}
In this representation, each subsystem experience its own emergent temporal flow, determined by the rate at which correlations with the clock evolve. Differences in gravitational potential, relative velocity, or cosmological embedding lead to distinct effective time parameters, even though all subsystems are correlated with the same underlying clock state. This structure provides a unified and internally consistent description of relative time dilation across multiple observers. Rather than postulating separate clocks or coordinate systems, temporal differences arise directly from the relational quantum state and the spacetime geometry in which each subsystem is embedded.
The framework therefore accommodates an arbitrary number of observers without introducing inconsistencies or preferred frames, and naturally reproduces known relativistic limits while remaining fundamentally quantum in origin.
\section{Physical Limits and Interpretation}
The unified emergent-time expression admits several physically relevant limiting cases. In the strong-gravity limit, such as near a black-hole horizon, the gravitational contribution dominates and the emergent temporal rate a approaches zero, reproducing extreme gravitational time dilation. In kinematic regimes involving relativistic motion, velocity-dependent terms govern the slowing of emergent time, consistent with special relativity.
In cosmological settings, accelerated expansion introduces an effective suppression of emergent temporal flow through horizon effects\cite{Mukhanov}, even for comoving observers. This reveals a close analogy between gravitational redshift at the level of emergent time, suggesting a common relational origin.
Across all these regimes, time dilation appears not as a modification of an external clock, but as a change in the rate at which quantum correlations evolve. This highlights the interpretational advantage of the emergent framework: relativistic effects are encoded directly into the structure of the quantum state rather than imposed geometrically by hand. 
\section{Conclusion and Outlook}
In this work, we have presented a unified framework in which time emerges as a relational quantum quantity derived from entanglement between physical subsystems and a global clock. The resulting emergent temporal flow is observer-dependent and naturally incorporates gravitational , kinematic, and cosmological effects.
massless or non-entangled entities, such as photon, do not experience emergent time \cite{CastroRuiz}, as they lack an internal relational clock. Strong gravitational fields produce the most pronounced suppression of temporal flow by reducing the rate of entanglement evolution, while cosmic acceleration induces subtler but conceptually similar effects through horizon dynamics.
Within this picture, the Schr\"odinger equation should be interpreted as an observer-dependent description of an underlying timeless quantum dynamics. When the emergent time parameter coincides with an intrinsic clock variable, standard quantum evolution is recovered. When it does not, the form of evolution changes while the fundamental dynamics remain intact.
The proposed framework bridges quantum mechanics and relativity at the conceptual level, providing a coherent foundations for understanding time as an emergent, relational, and fundamentally quantum phenomenon. Future work may explore extensions to interacting clocks, quantum gravitational back-reaction, and pre-geometric regimes.

\end{document}